\documentclass[preprint,aps]{revtex4}

\usepackage{graphicx}
\usepackage{dcolumn}
\usepackage{bm}

\begin{document}

\title{From normal state to magnetic storms in terms of fractal dynamics}

\author{Georgios Balasis}
\email{gbalasis@gfz-potsdam.de}
\affiliation{GeoForschungsZentrum Potsdam, Telegrafenberg, D-14473 Potsdam, 
Germany}
\author{Panayiotis Kapiris}
\author{Konstantinos Eftaxias}
\affiliation{Faculty of Physics, University of Athens, Panepistimiopolis, 
Zografos, 15784 Athens, Greece}

\date{\today} 

\begin{abstract}
We show that distinctive alterations in scaling parameters of $D_{st}$ index 
time series occur as a strong magnetic storm approaches. These alterations 
reveal a gradual reduction of complexity as the catastrophic event 
approaches. The increase of the susceptibility coupled with the transition 
from anti-persistent to persistent behavior may indicate that the onset of a 
severe magnetic storm is imminent. The preparation of a major magnetic storm 
could be studied in terms of ``Intermittent Criticality''. The analysis also 
suggests that the continuous scale invariance is partially broken into a 
discrete scale invariance symmetry.
\end{abstract}

\maketitle

Major magnetospheric disturbances are undoubtedly among the most important 
phenomena in space physics and also a core subject of space weather. 
They are relatively rare events: as in the case of atmospheric storms, 
earthquakes, solar flares, etc., the occurrence of geomagnetic storms rapidly 
decreases as their magnitude grows.

$D_{st}$ is the disturbance storm time index, computed from an average over 4 
magnetic observatories around the equator, and is considered to represent the
magnetospheric ring current contribution \cite{dag}. 
Since the development of a strong ring current is a defining feature of 
magnetic storms, the $D_{st}$ has been adopted as a proxy for magnetic storm 
severity \cite{lie}. 

In the context of complex systems self-organized criticality has been 
associated with natural hazards \cite{mal} 
such as earthquakes, landslides and forest-fires. The idea of using cellular 
automata to model the magnetospheric activity became popular as observations 
showed scale-invariant features. However, the debate between a forced 
(driven) and / or self-organized critical magnetosphere is not at all unique 
\cite{cpr,sit,fre}. 

The magnetosphere as a 
complex system manifests itself in linkages between space and time, producing 
scaling patterns and the emergence of fractal structures. We show that 
distinctive alterations in associated scaling parameters emerge (e.g., 
transition from anti-persistent to persistent behavior) as large 
magnetospheric disturbances approach 
(e.g., 6/11/2001, $D_{st} \simeq -300$ nT). 
The analysis also reveals the existence of discrete scale invariance, 
a system property invariant under only a discrete set of dilatations 
\cite{s98,s04}. 

\begin{figure}
\includegraphics[width=8.6cm]{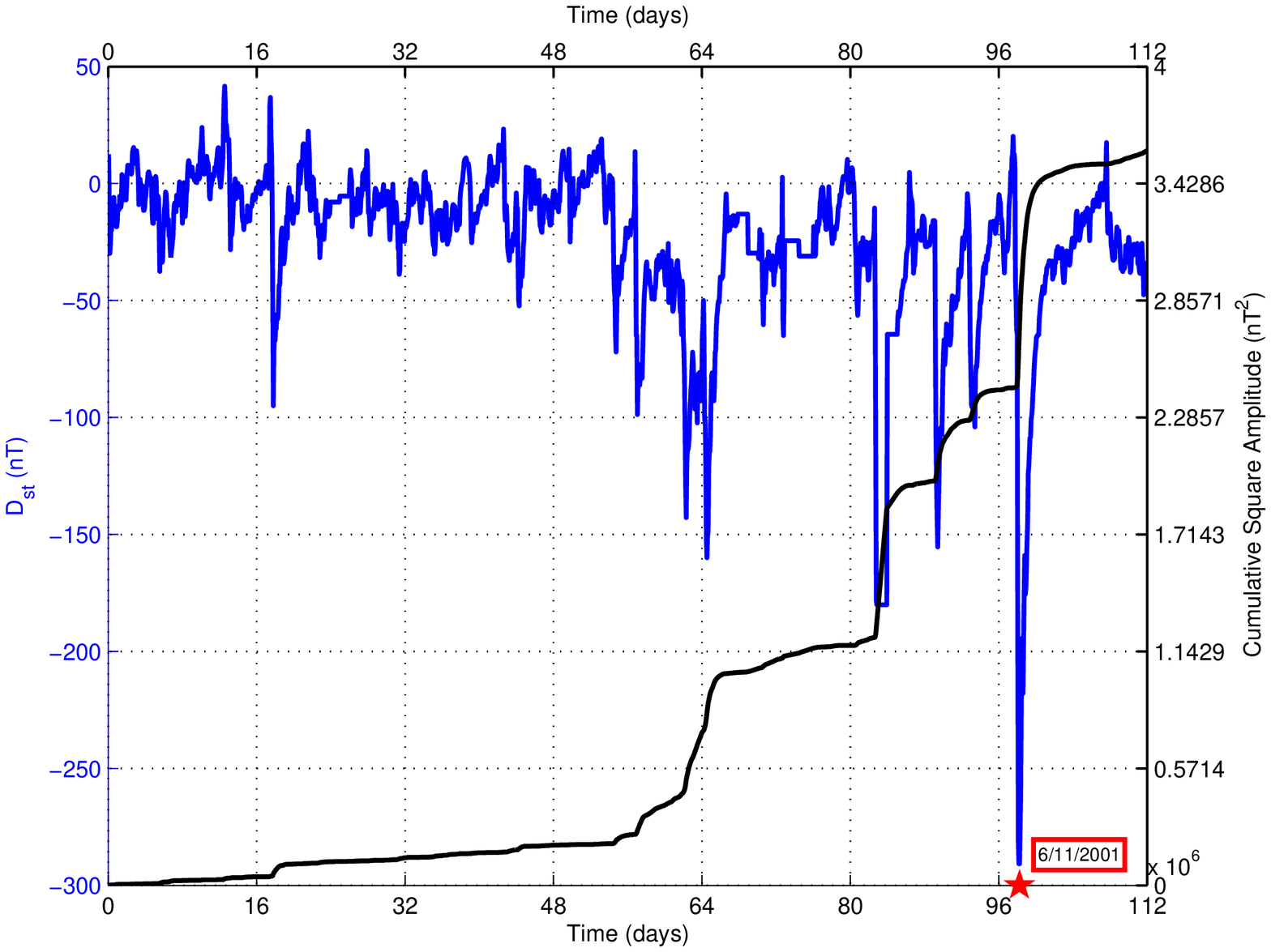}
\caption{\label{fig:1} $D_{st}$ time series and corresponding cumulative square amplitudes.
Red star denotes the peak of the magnetic storm of 6/11/2001 with 
$D_{st}$ = -292 nT.}
\end{figure}

\begin{figure}
\includegraphics[width=8.6cm]{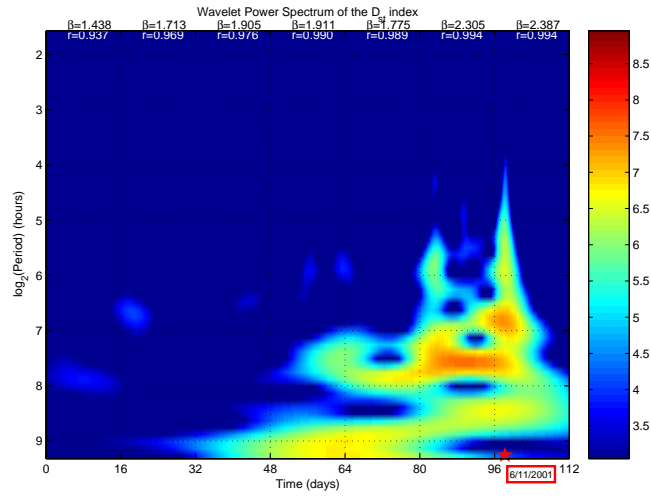}
\caption{\label{fig:2} Wavelet power spectrum of the $D_{st}$ time series, 
shown in Fig.~\ref{fig:1}. Linear correlation coefficients, $r$, and spectral 
exponents, $\beta$, calculated every 16 days are indicated on the top of each 
segment. Red star denotes the peak of the magnetic storm of 6/11/2001 with 
$D_{st}$ = -292 nT.}
\end{figure}

$D_{st}$ index time series were analyzed using a wavelet analysis technique 
\cite{bal}. In Fig.~\ref{fig:1}-\ref{fig:2}, the $D_{st}$ time series and the 
associated wavelet power spectrum are shown, respectively.

If a time series is a temporal fractal then a power-law of the form 
$S(f) \propto f^{-\beta}$ is obeyed. $S(f)$ is the power spectral density and 
$f$ is the frequency. The quantity $S(f)df$  may be understood as the 
contribution to the total power from those components of the time series, 
whose frequencies lie in the interval between $f$ and $f+df$. The spectral 
scaling exponent $\beta$ is a measure of the strength of time correlations. 
In a $\log S(f) - \log f$  representation the power spectrum is a line with 
slope $\beta$. The goodness of the fit of a time series to the power-law is 
assigned with the linear correlation coefficient, $r$, of this 
representation. 

A lot of work on complexity has been focused on power-laws, which describe 
the scaling properties of fractal processes and structures. Here, 
we examine whether distinctive alterations in the associated 
parameters, i.e., $r$ and $\beta$, emerge as a major magnetic storm 
approaches. For this purpose, the $D_{st}$ time variations were divided upon 
successive segments of 16 days and for each of these segments the 
parameters $r$ and $\beta$ were estimated and indicated at the top of the 
wavelet power spectrum in Fig.~\ref{fig:2}. (16 days is the shortest time 
window we can view here based on the index sampling rate and the definition 
of the wavelet transform.) 

The temporal evolution of $r$ (always above 0.94 and reaching values of 0.99 
at the end of the time interval considered here) means that the fit to the 
power-law is excellent. The fractal-law ($S(f) \propto f^{-\beta}$) observed 
indicates the existence of memory. This means that the current value of the 
geomagnetic signal is correlated not only with its most recent value but also 
with its long-term history in a scale-invariant, fractal manner, namely the 
system refers to its history in order to define its future. We observe a 
gradual increase of $r$ as the main event approaches. This suggests that the 
fractal character of the underlying processes and structures becomes clearer 
with time. 

The distribution of $\beta$ exponent is also shifted to higher values. This 
shift reveals several features of the underlying mechanism. As $\beta$ 
increases the spatial correlation in the time series also increases
\cite{tur}. 
This behavior indicates a gradual increase of the 
memory, and thus a gradual reduction of complexity in the underlying
dynamics. This suggests that the onset of a severe magnetic storm 
may represent a gradual transition from a less orderly state to a more 
orderly state (see also \cite{sit}). 

Maslov {\it et al.} \cite{mas} have formally established the relationship between 
spatial fractal behavior and long-range temporal correlations for a broad 
range of critical phenomena. By studying the time correlations in the local 
activity, they show that the temporal and spatial activity can be described 
as different cuts in the same underlying fractal. In a geometrical sense, 
$\beta$ specifies the strength of the signal's irregularity as well. The 
fractal dimension $D$ is calculated from the relation $D = (5 - \beta) / 2$
\cite{hen}, 
which, after considering the shift of $\beta$ to higher values, leads to a 
decrease of the fractal dimension as the magnetospheric crisis approaches. 
This may reflect that the action of anisotropy inherent to the system leads 
to the appearance of a clear preferred direction of elementary activities 
just before the main shock. Theoretical and experimental evidence support the 
former hypothesis: throughout the entire main and most of the early recovery 
phase of magnetic storms the geometry of the energy flow produces a highly 
asymmetric ring current configuration \cite{dag,lie}. 
The emergence of strong anisotropy 
rationalizes a further reduction of the complexity with time. 

The colour-type behavior of the power spectrum density ($\beta > 0$) means 
that the spectrum manifests more power at low frequencies than at high 
frequencies. The increase in the spectral exponent $\beta$ with time 
indicates the gradual enhancement of lower frequency fluctuations. This 
observation is consistent with the following physical picture: the activated 
substorms interact and coalesce to form larger fractal structures, i.e., the 
events are initiated at the lowest level of the hierarchy, with the smallest 
elements merging in turn to form larger and larger ones. This sign may be 
considered as candidate precursor of the forthcoming shock.

The $\beta$ exponent is related to the Hurst exponent, $H$, by the formula 
$\beta = 2 H +1$, with $0 < H < 1$ ($1 < \beta < 3$) for the fractional 
Brownian motion (fBm) random field model \cite {hen}. The exponent $H$ 
characterizes the persistent / anti-persistent properties of the signal 
\cite{k03}. 
The range $0 < H < 0.5$ ($1 < \beta < 2$) during the normal period (0 -- 80 
days) indicates anti-persistency, reflecting that if the fluctuations 
increase in a period, it is likely to decreasing in the interval immediately 
following and vice versa. Physically, this implies that fluctuations tend to 
induce stability within the system (negative feedback mechanism). The 
observed systematic increase of the $H$ ($\beta$) exponent during this stage 
indicates that the fluctuations become more correlated with time \cite{k04}. 

We pay attention to the fact that the time series appear persistent 
properties, $0.5 < H < 1$ ($2 < \beta < 3$), at 80 -- 112 days. This means 
that if the amplitude of fluctuations increases in a time interval it is 
likely to continue increasing in the interval immediately following. In other
words, the system tends toward irreversibility (positive feedback mechanism) 
\cite{k04}. 
$H = 0.5$ ($\beta = 2$) suggests no correlation between the repeated 
increments. Consequently, this particular value takes on a special physical 
meaning: it marks the transition between persistent and anti-persistent 
behavior in the time series. 

In Fig.~\ref{fig:1} the $D_{st}$ cumulative square amplitudes are also shown. A 
significant increase in the rate of energy release as the main geomagnetic 
storm approaches is observed. This may show that during the persistent epoch 
the system is not only near the peak of the magnetic storm in the sense of 
having power-law correlations, but also in terms of exhibiting high 
susceptibility.

Fractals have dimensions that are in general real numbers. The generalization 
from the set of integers to the set of real numbers embodies the transition 
from the symmetry of translation invariance to symmetry of scale invariance. 
Fractals are also described by fractal dimensions that belong to the complex 
numbers. In the context of critical phenomena, the complex fractal dimension 
is associated with a discrete scale invariance (DSI), i.e., to the invariance 
of the system or of its properties only under magnifications that are integer 
powers of a fundamental ratio. Interestingly, the appearance of DSI signifies 
a partial symmetry breaking of a continuous scale invariance and the 
emergence of DSI on characteristic scales. 

DSI manifests itself in data by log-periodic corrections to scaling 
\cite{s98,s04,hua}. 
The typical formula for log-periodicity in time is given by 
$E(t)=A+B(t_f-t)^m\{1+Ccos[\omega log(t_f-t)+\phi]\}$, where $E (t) $ is the 
cumulative energy released, $t_f$ is the time of the main shock (storm peak), 
$\omega$ is the frequency and $\phi$ is just an offset. 
	
\begin{figure}
\includegraphics[width=8.6cm]{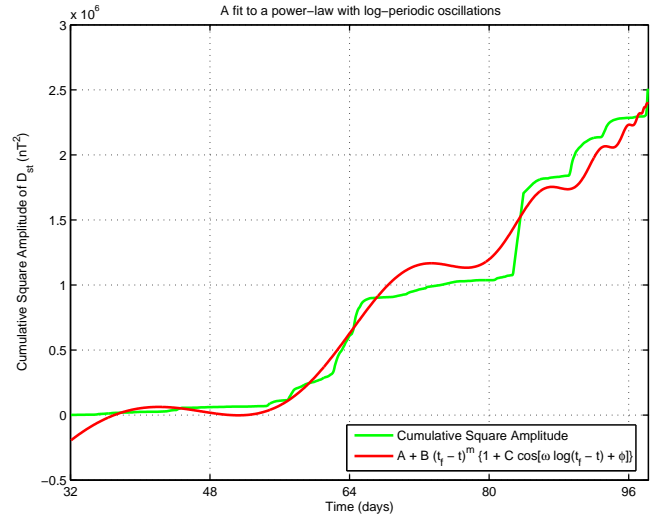}
\caption{\label{fig:3} A fit of the cumulative square amplitudes, shown in 
Fig.~\ref{fig:1}, to a power-law with log-periodic oscillations.}
\end{figure}

We focus on Fig.~\ref{fig:1} and in the energy oscillations observed prior to the 
main magnetic event (32 -- 96 days).  One can observe a trace of oscillations 
modulating the main power-law behavior in the energy density. In
Fig.~\ref{fig:3} we 
fit a power-law with log-periodic oscillations to the cumulative square 
$D_{st}$ amplitudes. It is clear that a law of this form can adequately 
describe the observations.

As expected, the log-periodic oscillations are modulated in frequency with a 
geometric increase of the frequency on the approach to the time $t_f$: the 
intermittent accelerations and quiescences of geomagnetic activity around the 
power-law acceleration become more closely spaced as the main event is 
approached. The aforementioned behavior reflects a preparation stage for 
major geomagnetic storms, in which pre-storm activities occur at particular 
discrete times and not in a continuous fashion: these discontinuities in turn 
mirror the localized and threshold nature of the underlying mechanism. It is 
this ``punctuated'' physics which gives rise to the scaling precursors 
modeled mathematically by the log-periodic correction to scaling. 

By monitoring the temporal evolution of the fractal spectral characteristics 
in $D_{st}$ we find that distinctive alterations 
in the associated scaling parameters indicate a transition from the normal 
state to an abnormal state (major magnetic storm) as following: (i) 
Emergence of long-range correlations, i.e., appearance of  memory effects. 
This implies a multi-time-scale cooperative activity of numerous activated 
geomagnetic events. (ii) Increase of the spatial correlation in the time 
series with time. This indicates a gradual transition from a less orderly 
state to a more orderly state. (iii) Decrease of the fractal dimension of the 
variations with time, i.e., appearance of strong anisotropy in elementary 
activities. (iv) Existence of strong anti-persistent behavior in the first 
epoch of the geomagnetic activity, i.e., prior to the severe 
magnetic storm. (v) Decrease of the anti-persistent behavior with time. (vi) 
Emergence of persistent properties in the ``tail'' of the time series. (vii) 
Predominance of large geomagnetic events with time. (viii) Significant 
acceleration of the energy release as the main shock approaches, i.e., 
increase of the susceptibility of the system. (ix) Gradual appearance of 
higher frequencies in the spectrum with simultaneous increase of the 
amplitudes at each emission rate as the magnetic storm peak approaches, 
mainly characterizing lower emission rates. 

A question that arises is whether the evolution towards global 
instability is inevitable after the appearance of distinctive symptoms in the 
geomagnetic variations. The emergence of persistent behavior, the increase of 
the susceptibility of the system, the predominance of large geomagnetic 
events, the coherent fluctuations at all scales, may indicate that the 
generation of a very strong magnetic storm becomes, indeed, unavoidable.

The aforementioned crucial footprints (including temporal alterations in 
associated scaling parameters) distinguish the dynamics of a complex system 
close to its final instability. These may indicate the following scenario 
for the generation of a severe magnetic storm. During the normal 
period, the system is in a 
anti-persistent state, with a restricted and systematically fluctuating 
correlation length. Long-range correlations gradually build up through local 
interactions until they extend throughout the entire system. The smaller 
geomagnetic events are the agents by which longer correlations are 
established. A population of small events will advance the correlation length 
by an amount depending on its magnitude and magnetosphere state, triggering 
very strong magnetic storms only if the condition is right: in a 
anti-persistent regime a population of small events leads to a decaying 
activity, always dying out. In the 
persistent state is just able to continue ``indefinitely''. This explains in 
a natural way why not every geomagnetic event can induce geomagnetic 
activity. A large geomagnetic event destroys long correlations on 
its associated network, creating a new normal period during which the process 
repeats by rebuilding correlation lengths towards the next large event. Thus 
a large shock may act as a sort of ``critical point'' dividing the magnetic 
storm into a period of growing correlations before the great event and a 
relatively uncorrelated phase after. 

The aforementioned evolution may be overall characterized as ``Intermittent 
Criticality'' \cite{bow} 
that predicts a time-dependent variation in 
the activity as the ``critical point'' is approached, implying, in contrast 
to self-organized criticality (SOC), a degree of predictability. 

The analogies with the dynamics of the SOC model for the magnetosphere have 
been realized by numerous authors \cite{sit,fre}. 
Characteristically, the scale-free structure of the aurora is argued to come 
from a scale-free structure of a SOC magnetosphere \cite{fre}. 
However, numerous 
authors are cautious with this suggestion. A relevant question, emphasized by 
Consolini and Chang  \cite{con}, is how well the assumption of the SOC model 
is met in magnetosphere. We also recall the debate between a driven and an 
internal origin for intermittent scale-free dynamics in magnetosphere. Work 
continues on this issue, which is an example of a generic problem of complex 
systems coupled to complex drivers \cite{fre}. 
We think that the results of this study may help to approach the real 
dynamics of magnetosphere. In any case the present study suggests that it can 
be important to distinguish between SOC and intermittent criticality in the 
study of the magnetic storm cycle. A proper recognition and understanding of 
tuning parameters may lead to the development of improved magnetospheric 
models having higher performance reliability. One of the main features in the 
complexity is the role that the topological disorder plays in such systems. 
The range of size scales characterizing heterogeneities of the thresholds 
might be acted as a tuning parameter of the underlying final magnetic storm 
dynamics. 

We bear in mind that our analysis reveals an interesting transition from the 
anti-persistent to the persistent regime. The anti-persistent behavior 
characterizes the magnetosphere during substorms, while the catastrophic 
events are in reasonable agreement with persistent models. We note that 
Sitnov {\it et al.} \cite{sit} have also suggested two differents regimes: while 
the substorm activity resembles second-order phase transitions, the largest 
substorms avalanches are shown to reveal the features of first-order 
non-equilibrium transitions. The corresponding pictures in each regime are 
not in contradiction. 

It is important to stress the practical consequence of log-periodic 
structures. For forecasting purpose, it is much more constrained and thus 
reliable to fit a part of an oscillating data than a simple power-law which 
can be quite degenerate especially in the presence of noise. This idea has 
been noted and is vigorously investigated in several applied domains, such as 
earthquakes, rupture and financial crashes \cite{s04}. 


\begin{acknowledgments}
This research was supported by DFG's research grant MA 2117/3, as part of
the Priority Program SPP 1097. Helpful discussions with P. Bedrosian, H.
L\"uhr and S. Maus are gratefully acknowledged.
\end{acknowledgments}


\newpage

\end{document}